\documentclass{aa}
\usepackage{graphicx}
\usepackage{txfonts}

\begin{document} 

\title{The sulfur depletion problem: 
upper limits on the H$_{2}$S$_{2}$, HS$_{2}^{.}$, and S$_{2}$ gas-phase abundances toward the low-mass warm core IRAS16293-2422\thanks{
The reduced spectra for the observed 16 GHz at the original spectral resolution are available in electronic form at the CDS via anonymous 
ftp to cdsarc.u-strasbg.fr (13.79.128.5) or via http://cdsweb.u-strasbg.fr/cgi-bin/qcat?J/A+A}}
\titlerunning{Upper limits on H$_{2}$S$_{2}$, HS$_{2}^{.}$, and S$_{2}$ abundances toward IRAS16293-2422}
\author{R. Mart\'in-Dom\'enech \inst{\ref{inst1}} \and I. Jim\'enez-Serra \inst{\ref{inst2},\ref{inst3}} \and G.~M. Mu\~noz Caro \inst{\ref{inst1}} 
\and H.~S.~P. M\"uller \inst{\ref{inst4}} \and A. Occhiogrosso \inst{\ref{inst2}} \and L. Testi \inst{\ref{inst3}} \and P.~M. Woods \inst{\ref{inst5}} 
\and S. Viti \inst{\ref{inst2}}}

\institute{Centro de Astrobiolog\'ia (INTA-CSIC), Ctra. de Ajalvir, km 4, Torrej\'on de Ardoz, 28850 Madrid, Spain
\email{rmartin@cab.inta-csic.es}\label{inst1}
\and
University College London, 132 Hampstead Road, London NW1 2PS, United Kingdom\label{inst2}
\and
European Southern Observatory, Karl-Schwarzschild-Str. 2, 85748 Garching, Germany\label{inst3}
\and
I.~Physikalisches Institut, Universit\"at zu K\"oln, Z\"ulpicher. Str. 77, 50937 K\"oln, Germany\label{inst4}
\and 
Astrophysics Research Centre, Dept. of Mathematics \& Physics, Queen's University Belfast, Belfast BT7 1NN, UK\label{inst5}
}

\date{}

\abstract
{A fraction of the missing sulfur in dense clouds and circumstellar regions could be in the form 
of three species not yet detected in the interstellar medium: H$_{2}$S$_{2}$, HS$_{2}^{.}$, and S$_{2}$ 
according to experimental simulations performed under astrophysically relevant conditions. 
These S-S bonded molecules can be formed by the energetic processing of H$_{2}$S-bearing ice mantles on dust grains, 
and subsequently desorb to the gas phase.}
{The detection of these species could partially solve the sulfur depletion problem, 
and would help to improve our knowledge of the poorly known chemistry of sulfur in the interstellar medium. 
To this purpose 
we calculated the frequencies and expected intensities of the rotational transitions not previously reported, 
and performed dedicated ground-based observations toward the low-mass warm core IRAS16293-2422, 
a region with one of the highest measured gas-phase H$_{2}$S abundances.}
{Observations in the submillimeter regime were obtained with the APEX 12 m telescope during 15 hours of observation.
A total of $\sim$16 GHz were covered in a range of about 100 GHz, 
targeting a wide selection of the predicted rotational transitions of the three molecules.} 
{
The 1$\sigma$ noise rms values were extracted in the spectral regions where the targeted species should have been detected.  
These values were a factor of 2 - 7 lower than those reached by previous observations toward the same source, 
and allowed us to estimate a 1$\sigma$ upper limit to their molecular abundances of 
$\leq$8.1 x 10$^{-9}$, $\leq$1.1 x 10$^{-8}$, and $\leq$2.9 x 10$^{-7}$ relative to H$_{2}$, for H$_{2}$S$_{2}$, HS$_{2}^{.}$, and S$_{2}$, 
respectively.} 
{The upper limit abundances of the three molecules containing the S$_{2}$ unit are up to two orders of magnitude lower than the H$_{2}$S abundance 
in the source, 
and one order of magnitude lower than the expected abundances from the experimental simulations using ice analogs. 
Subsequent gas-phase chemistry after desorption could lower the abundances of the three species to undetectable levels in our observations.} 


\maketitle

\section{Introduction}
\label{intro}

Sulfur, the tenth most abundant element in the Galaxy, is of particular interest from an astrochemical point of view. 
On the one hand it has been suggested that sulfur-bearing species can act as chemical clocks in star forming regions 
\citep{charnley97,hatchell98,viti01,wakelam04a,wakelam11}. 
On the other hand, along with carbon, hydrogen, nitrogen, oxygen and phosphorus, it is one of the elements commonly present 
in molecules of biotic interest.  

Previous observations have shown 
that sulfur is depleted in dense clouds and circumstellar regions around young stellar objects (YSOs) 
The S-bearing species already detected in these regions account for only $\sim$ 0.1\% of
its estimated cosmic abundance \citep[1.23 x 10$^{-5}$ N$_{H}$;][]{tieftrunk94}.   
It has been proposed 
that the missing sulfur is locked onto the icy mantles of dust grains \citep[e.g.,][]{millar90,jansen95,ruffle99}. 
Until now, only OCS \citep{geballe85,palumbo95} 
and SO$_{2}$ \citep{boogert97} 
have been firmly detected in icy grain mantles toward high-mass protostars, 
but their estimated abundances 
are on the order of $\sim$ 0.5\%, and $\sim$ 0.8 - 4.0\%, respectively, of the total cosmic S abundance \citep{boogert97}. 
%

In cometary ices H$_{2}$S is the most abundant S-bearing molecule, with an abundance of up to 1.5\% relative to water 
\citep{bockelee00}, 
or 1.5$\times$ 10$^{-6}$ relative to H$_{2}$. 
The presence of H$_{2}$S 
in interstellar ices 
was inferred from IR observations toward the high-mass protostar W33A \citep{geballe85}.  
However, 
a robust detection of this molecule in interstellar ices has not been reported yet,  
probably because of the overlapping of the 3.92 $\mu m$ IR band of H$_{2}$S with a strong IR feature of methanol.   
The upper limits found for these species toward dense clouds and circumstellar regions 
\citep[0.6 - 1.6 $\times$ 10$^{-6}$, and 0.04 - 0.12 $\times$ 10$^{-6}$ relative to H$_{2}$, respectively; ][]{smith91} 
account for only 10\% of the cosmic S abundance. 

A fraction of the H$_{2}$S present in interstellar ices could be energetically processed 
by UV-photons, X-rays, or cosmic rays in dense cloud interiors and regions around YSOs, leading to the formation of other S-bearing species.  
These new molecules would subsequently be released to the gas phase by means of 
photon or ion-induced desorption, and also thermal desorption 
in the hot regions of  
the circumstellar envelopes around YSOs, 
harboring part of the missing sulfur. 
This is indeed a very plausible scenario, since several of the gas-phase S-containing molecules already observed in hot cores, 
such as H$_{2}$S, SO$_{2}$, OCS, SO, H$_{2}$CS, HCS$^{+}$, and NS \citep{vandertak03,izaskun12}, 
have abundances that cannot be explained with gas-phase-only chemical models \citep{doty04,viti04,wakelam04a}. 
This scenario has been tested in the laboratory for the past thirty years.
Experimental simulations of the irradiation of interstellar ices containing H$_{2}$S under astrophysically relevant conditions have been 
performed  
using 
UV photons \citep{grim87,jimenez11,jimenez14}, 
X-rays \citep{jimenez12}, 
or ions \citep{moore07,ferrante08,garozzo10}. 
Energetic processing of H$_{2}$S-bearing ices readily generates sulfur-sulfur bonds, and 
the main S-bearing products in these experiments are usually H$_{2}$S$_{2}$ and HS$_{2}^{.}$. 
The molecule H$_{2}$S$_{2}$ can subsequently photodissociate forming S$_{2}$ and S$_{3}$, which 
have recently been detected by \citet{jimenez11}. 
These molecules with two S atoms could thus contain a significant fraction of the missing sulfur in dense clouds and circumstellar regions, 
but their observation in the solid or gas phase have not yet been reported. 

Dense circumstellar envelopes around YSOs such as high-mass hot cores and low-mass warm cores (or hot corinos) 
are prime candidates for the detection of these S-bearing molecules in the gas phase, 
since ice mantles are largely affected by intense UV and/or X-ray irradiation from their central protostars 
before their thermal desorption.  
The low-mass warm core IRAS16293-2422 (or hot corino) is known to show an active chemistry as demonstrated by its rich molecular line spectrum 
\citep{caux11,jorgensen11}. 
The derived gas-phase H$_{2}$S abundance toward this object is one of the highest measured in low-mass warm cores 
\citep[5 x 10$^{-7}$;][]{wakelam04b} 
and is therefore a good candidate to test whether an important fraction of the missing sulfur is contained in the form of 
any of these S-bearing species coming from H$_{2}$S. 
IRAS16293-2422 is a binary system whose sources A and B are spatially separated by 5" \citep{mundy92}. 
Among the two sources, IRAS16293-2422 A shows a very rich sulfur chemistry with several sulfur-bearing molecules detected, 
such as H$_{2}$S, SO, SO$_{2}$, HCS$^{+}$, H$_{2}$CS, CS, and OCS \citep{wakelam04b,caux11,jorgensen11}. 
The measured rotational temperature of the hot corino is T$_{\rm{rot}}$ = 100 K \citep{ceccarelli00}. 


We present the first single-dish observations targeting 
a wide selection of the predicted rotational transitions for 
three of the products detected in the experimental simulations 
(H$_{2}$S$_{2}$, HS$_{2}^{.}$, and S$_{2}$) toward IRAS16293-2422. 
Although none of the three species was firmly detected,  
our observations
 provide stringent upper limits to their molecular abundances,  
constraining the efficiency of UV-photoformation of H$_{2}$S$_{2}$, HS$_{2}^{.}$, and S$_{2}$ in interstellar ices. 
This paper is organized as follows. 
Section \ref{lab-spec} presents spectroscopic information of the three targeted molecules. 
In particular, the data provided for H$_{2}$S$_{2}$ was not previously documented.  
Section \ref{obs} describes the observations that have been performed, 
while the results are shown in Section \ref{results}. 
The upper limits derived from our observations are discussed in Section \ref{upper}. 
In Section \ref{comparison} these upper limits are compared to the expected abundances from the experimental simulations with ice analogs,  
and in Section \ref{model} to the estimated abundances from theoretical simulations taking into account the subsequent gas-phase chemistry.  
The conclusions are summarized in Section \ref{conclusiones}.

\section{Laboratory spectroscopy of the sulfur-containing molecules in the gas phase}
\label{lab-spec}

Disulfane, H$_{2}$S$_{2}$, also known as hydrogen disulfide, is a heavy homologue of hydrogen 
peroxide, H$_{2}$O$_{2}$. Whereas H$_{2}$O$_{2}$ displays large torsional splitting in the ground 
vibrational state, this splitting is hard to resolve for H$_{2}$S$_{2}$ even in the laboratory. 
It is an asymmetric top rotor very close to the prolate limit, i.e., its $A$ rotational constant 
of almost 147~GHz is much larger than $B$ and $C$, which are almost equal and slightly smaller 
than 6.97~GHz. Its dipole moment of about 1.1~D \citep{many_calculated_dipole_moments_2009} 
is along the $c$-axis. The two equivalent H nuclei result in a 3~:~1 spin-statistical ratio 
between \textit{ortho} and \textit{para} levels which are described by $K_c$ being odd and 
even, respectively. A large body of transition frequencies up to 421~GHz were taken from 
\citet{H2S2_rot_1990} with additional terahertz data from \citet{H2S2_rot_1994} and \citet{H2S2_rot_1996}. 
The precision of the data appeared to be 20 or 30~kHz for the most part. 
The $K = 3 - 2$ data in \citet{H2S2_rot_1994} is too low compared with the data reported in \citet{H2S2_rot_1990} by about 300~kHz with rather 
small scatter.  
The absence of a reference signal in, e.g., \citet{H2S2_rot_1994} could be an 
explanation, see SO data in \citet{SO_rot_Koeln_1994} and \citet{SO_rot_Koeln_1996}. Therefore, the 
$K = 3 - 2$ data from \citet{H2S2_rot_1994} were omitted from the fit. As a consequence, 
predicted transition frequencies should be viewed with some caution. 
As the result of this work,  
a catalog entry was 
created for the Cologne Database for Molecular Spectroscopy, CDMS, \citep{CDMS_2001,CDMS_2005} 
with the spectroscopic information of this species, including predicted frequencies and intensities for its rotational transitions.

Elimination of one H atom from disulfane 
leads to the 2A$^{''}$ thiosulfeno radical, HS$_{2}^{.}$. 
The $A$ rotational constant is roughly twice that of H$_{2}$S$_{2}$, while $B$ and $C$ are 
about 1~GHz larger. The dipole moment components were calculated as $\mu_a = 1.16$~D and 
$\mu_b = 0.83$~D \citep{HS2_ab_initio_2008}. Predictions of the HS$_{2}^{.}$ rotational spectrum 
were also taken from the CDMS. They were based on the analysis of \citet{HS2_rot_2000} 
which included lower frequency data from \citet{HS2_rot_1994}. $^1H$ hyperfine structure 
is usually small and was often not resolved in the laboratory. Electron spin-rotation 
splitting is small for $a$-type transitions, e.g., on the order of 200~MHz and 1~GHz for 
$K_a = 0 - 0$ and $1 - 1$, respectively. It is several tens of gigahertz for $b$-type 
transitions.

Elimination of the second H atom yields S$_{2}$, the smallest molecule containing only sulfur. 
It is a heavy homologue of O$_{2}$, the ground electronic state is $^3\Sigma _g ^-$. Because 
of its symmetry, it does not have an electric dipole moment, but it has a magnetic dipole moment 
which gives rise to a weak rotational spectrum. The selection rules are $\Delta N = 0$, $\pm2$ 
and $\Delta J = 0$, $\pm1$. The predictions were taken from the JPL catalog \citep{JPL-catalog_1998}. 
They are based on rotational data from \citet{S2_rot_1979} with additional ground state 
combination differences taken from the $b - X$ electronic spectrum \citep{S2_b-X_1986}. 
The magnetic $g$-factors were redetermined by \citet{S2_rot_1979}. The rotational spectrum 
of O$_{2}$ is quite close to Hund's case (b), in which the three $\Delta N = 2$ transitions 
occur quite close in frequency. The rotational splitting in S$_{2}$ is considerably smaller 
than in O$_{2}$ whereas the splitting caused mainly by the spin-spin coupling is much larger. 
As a consequence, the rotational spectrum is much more complex in its structure, closer 
to Hund's case (a) at low rotational quantum numbers $N$ and closer to Hund's case (b) at 
higher $N$. Relatively strong transitions are those with $\Delta J = 0$, those with 
$\Delta N = 0$ are also comparatively strong.

\begin{table*}
\centering
\caption{Targeted transitions and observational results}
\begin{tabular}{ccccccccc}
\hline
\hline
& & Frequency & E$_{\rm{up}}$ & g$_{\rm{up}}$ & A$_{\rm{ul}}$ & beam size& rms $^{(1)}$ & N\\
Molecule & Transition & (MHz) & (K) & & (s$^{-1}$) & ($\arcsec$) & (K) & (cm$^{-2}$)\\
\hline
H$_{2}$S$_{2}$ & 6$_{1,5}$-5$_{0,5}$ & \bf{223535.85} & 20.8 & 39 & 4.23 x 10 $^{-5}$ & 28 & ... & ...$^{(2)}$\\
H$_{2}$S$_{2}$ & 13$_{2,11}$-14$_{1,13}$ & 224367.15 & 87.8 & 81 & 1.64 x 10 $^{-5}$ & 28 & 0.0018 & $\leq$ 1.7 x 10$^{15}$\\
H$_{2}$S$_{2}$ & 13$_{2,12}$-14$_{1,14}$ & 224657.47 & 87.8 & 27 & 1.65 x 10 $^{-5}$ & 28 & ... & ...$^{(2)}$\\
H$_{2}$S$_{2}$ & 8$_{1,7}$-7$_{0,7}$ & \bf{251423.70} & 30.8 & 51 & 5.92 x 10 $^{-5}$ & 25 & 0.0026 & $\leq$ 6.1 x 10$^{14}$\\
H$_{2}$S$_{2}$ & 11$_{2,9}$-12$_{1,11}$ & 252271.50 & 71.0 & 69 & 2.25 x 10 $^{-5}$ & 25 & 0.0025 & $\leq$ 1.7 x 10$^{15}$\\
H$_{2}$S$_{2}$ & 11$_{2,10}$-12$_{1,12}$ & 252486.63 & 71.0 & 23 & 2.25 x 10 $^{-5}$ & 25 & ... & ...$^{(2)}$\\
H$_{2}$S$_{2}$ & 10$_{1,9}$-9$_{0,9}$ & \bf{279312.25} & 43.5 & 63 & 8.04 x 10 $^{-5}$ & 22 & ... & ...$^{(2)}$\\
H$_{2}$S$_{2}$ & 9$_{2,7}$-10$_{1,9}$ & 280174.30 & 57.0 & 57 & 2.94 x 10 $^{-5}$ & 22 & ... & ...$^{(2)}$\\
H$_{2}$S$_{2}$ & 9$_{2,8}$-10$_{1,10}$ & 280325.68 & 57.0 & 19 & 2.94 x 10 $^{-5}$ & 22 & 0.0022 & $\leq$ 3.7 x 10$^{15}$\\
H$_{2}$S$_{2}$ & 14$_{1,13}$-13$_{0,13}$ & \bf{335087.32} & 77.0 & 87 & 1.37 x 10 $^{-4}$ & 19 & ... & ...$^{(2)}$\\
H$_{2}$S$_{2}$ & 5$_{2,3}$-6$_{1,5}$ & 335971.39 & 36.9 & 33 & 4.05 x 10 $^{-5}$ & 19 & 0.0036 & $\leq$ 2.1 x 10$^{15}$\\
H$_{2}$S$_{2}$ & 5$_{2,4}$-6$_{1,6}$ & 336029.02 & 36.9 & 11 & 4.05 x 10 $^{-5}$ & 19 & ... & ...$^{(2)}$\\
\hline
HS$_{2}^{.}$ & 16$_{0,16,16,15}$-15$_{0,15,15,14}$ $^{(3)}$ & 252270.66 & 103.1 & 31 & 1.22 x 10 $^{-4}$ & & & \\[-1ex]
HS$_{2}^{.}$ & 16$_{0,16,16,16}$-15$_{0,15,15,15}$ $^{(3)}$ & 252270.71 & 103.1 & 33 & 1.22 x 10 $^{-4}$ & \raisebox{1.5ex}{25} & \raisebox{1.5ex}{0.0028} 
& \raisebox{1.5ex}{$\leq$ 8.0 x 10$^{14}$}\\
HS$_{2}^{.}$ & 4$_{1,3,5,4}$-4$_{0,4,5,4}$ $^{(3)}$ & 280022.63 & 21.0 & 9 & 7.45 x 10 $^{-5}$ & & & \\[-1ex]
HS$_{2}^{.}$ & 4$_{1,3,5,5}$-4$_{0,4,5,5}$ $^{(3)}$ & 280023.28 & 21.0 & 11 & 7.49 x 10 $^{-5}$ & \raisebox{1.5ex}{$22$} & \raisebox{1.5ex}{$0.0027$} 
& \raisebox{1.5ex}{$\leq$ $1.6$ x $10^{15}$} \\
HS$_{2}^{.}$ & 21$_{1,20,22,22}$-20$_{1,19,21,21}$ $^{(3)}$ & 333247.58 & 189.7 & 45 & 2.84 x 10 $^{-4}$ & & & \\[-1ex]
HS$_{2}^{.}$ & 21$_{1,20,22,21}$-20$_{1,19,21,20}$ $^{(3)}$ & 333247.67 & 189.7 & 43 & 2.84 x 10 $^{-4}$ & \raisebox{1.5ex}{19} & \raisebox{1.5ex}{0.0052} 
& \raisebox{1.5ex}{$\leq$ 1.3 x 10$^{15}$}\\
\hline
S$_{2}$ & 11$_{10}$-9$_{9}$ & 224301.13 & 82.1 & 21 & 3.02 x 10 $^{-7}$ & 28 & 0.0017 & $\leq$ 2.2 x 10$^{16}$\\
S$_{2}$ & 15$_{14}$-13$_{13}$ & 333685.77 & 126.3 & 29 & 7.78 x 10 $^{-7}$ & 19 & 0.0039 & $\leq$ 2.3 x 10$^{16}$\\
\hline
\end{tabular}
\begin{list}{}
\item(1) rms is given in 5 km s$^{-1}$ bins, 
\item(2) since these spectral regions present line contributions from other species 
near the frequency of the H$_{2}$S$_{2}$ transition, 
we have not used the corresponding noise rms value to compute the 1$\sigma$ upper limits to the column density, 
\item(3) the last two quantum numbers designate the spin quanta.  
\end{list}
\label{lineas}
\end{table*}

\section{Observations}
\label{obs}

The observations were carried out with the APEX 12 m telescope located at the high altitude site of Llano Chajnantor (Chile) in 2013, between 
August 30 and September 10, under good (PWV = 1.1 - 2.2 mm) weather conditions. 
We used the Swedish Heterodyne Facility Instruments (SHeFI; Vassilev et al. 2008) APEX-1 and APEX-2 as frontends 
with four different frequency setups of 4 GHz bandwidth each, 
covering a total of $\sim$16 GHz in a range of about 100 GHz.  
The eXtended bandwith Fast Fourier Transform Spectrometer \citep[XFFTS;][]{klein12} 
was used as backend. 
The XFFTS yields a spectral resolution of 76 KHz, which corresponds to $\sim$ 0.07 - 0.10 km s$^{-1}$ in the spectral range observed. 

The position-switching mode was chosen to perform the on/off observations.
The on position was centered toward IRAS16293-2422 A, $\alpha_{J2000}$ = 16$^{h}$32$^{min}$22.9$^{s}$, 
$\delta_{J2000}$ = -24$^{\circ}$28$\arcmin$37.0$\arcsec$, 
while the off position was located at $\alpha_{J2000}$ = 16$^{h}$32$^{min}$09.4$^{s}$, 
$\delta_{J2000}$ = -24$^{\circ}$28$\arcmin$33.0$\arcsec$, which is free of any C$^{18}$O and $^{13}$CO emission \citep{wakelam04b}. 
The half-power beamwidth of the telescope is $\sim$ 30 - 25$\arcsec$ for the APEX-1 receiver, and $\sim$ 23 - 17$\arcsec$ for the APEX-2 receiver. 
Therefore, emission from sources A and B of the binary protostellar system IRAS16293-2422 could not be spatially resolved. 
The receivers were tuned to single sideband, and the intensities were measured in units of T$_{\rm{A}}^{*}$. 
The conversion to T$_{\rm{mb}}$ was done by using main beam efficiencies of 0.75 and 0.73 for APEX-1 and APEX-2, respectively. 
Calibration and reduction of the data was previously performed with the software CLASS of the GILDAS package. 

Our observations covered a total of 17 H$_{2}$S$_{2}$ transitions, 96 HS$_{2}^{.}$ transitions, and 6 S$_{2}$ transitions. 
For H$_{2}$S$_{2}$, we selected the transitions with low upper level energy (a total of three transitions per frequency setup). 
The central frequency of the four observational setups corresponds to the frequencies of the selected H$_{2}$S$_{2}$ transitions 
expected to be the most intense in each setup, and are boldfaced in Table \ref{lineas}. 
For HS$_{2}^{.}$ and S$_{2}$, 
we selected only 
those transitions with low upper level energy that fell in a clean region of the spectrum.
The selected lines for H$_{2}$S$_{2}$, HS$_{2}^{.}$, and S$_{2}$ are listed in Table \ref{lineas}.

\section{Results}
\label{results}

Our observations toward the IRAS16293-2422 hot corino reached rms values which are a factor 2 - 7 better than those reached by the 
TIMASS survey \citep{caux11} in the 197 GHz - 280 GHz frequency range, while they are comparable in the 328 GHz - 366 GHz range. 
%
For the analysis of the spectra, 
we resampled the observations to a resolution of 0.5 km s$^{-1}$, 
and assumed a central radial velocity of v$_{\rm{lsr}}$ = 3.2 km s$^{-1}$ for IRAS16293 A \citep{jorgensen11}. 

Lines detected at frequencies near those of the targeted transitions in Table \ref{lineas} were assigned to previously detected species. 
The observed lines were then fitted with Gaussians (when possible, more than one Gaussian was used in the case of blended lines). 
For every putative assigned line, we used the integrated intensity extracted from the fit, and a  
simple model assuming local thermodinamical equilibrium to 
readily  
extract column densities 
that were  
compared 
with the column densities derived in previous works for the same species. 
This way, we could confirm the quality of our observations, and even 
assign lines 
corresponding to previously detected species 
that were not documented in previous works (see Section \ref{previas}).  

Putative assignments of transitions corresponding to the targeted species were, however, ruled out 
because of missing lines (see Section \ref{azufradas}).

\begin{figure*}
\centering
\includegraphics[width=18cm]{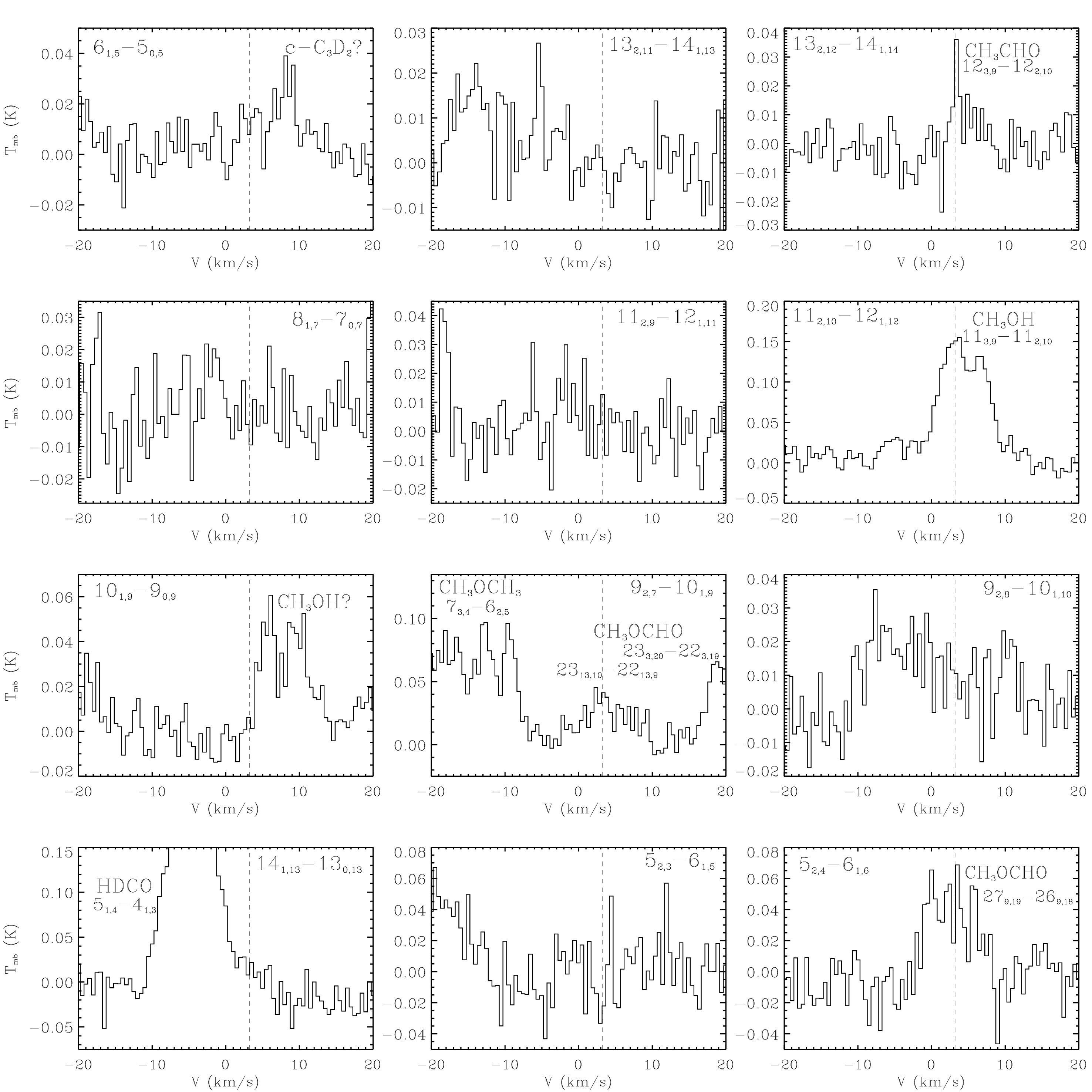}
\caption{Spectral windows where the H$_{2}$S$_{2}$ transitions listed in Table \ref{lineas} should be detected. 
Vertical dashed lines mark a central radial velocity of v$_{\rm{lsr}}$ = 3.2 km s$^{-1}$, assumed for IRAS16293 A \citep{jorgensen11}. 
}
\label{h2s2}
\end{figure*}

\subsection{Detection of species not containing sulfur}
\label{previas}

\begin{figure*}
\centering
\includegraphics[width=18cm]{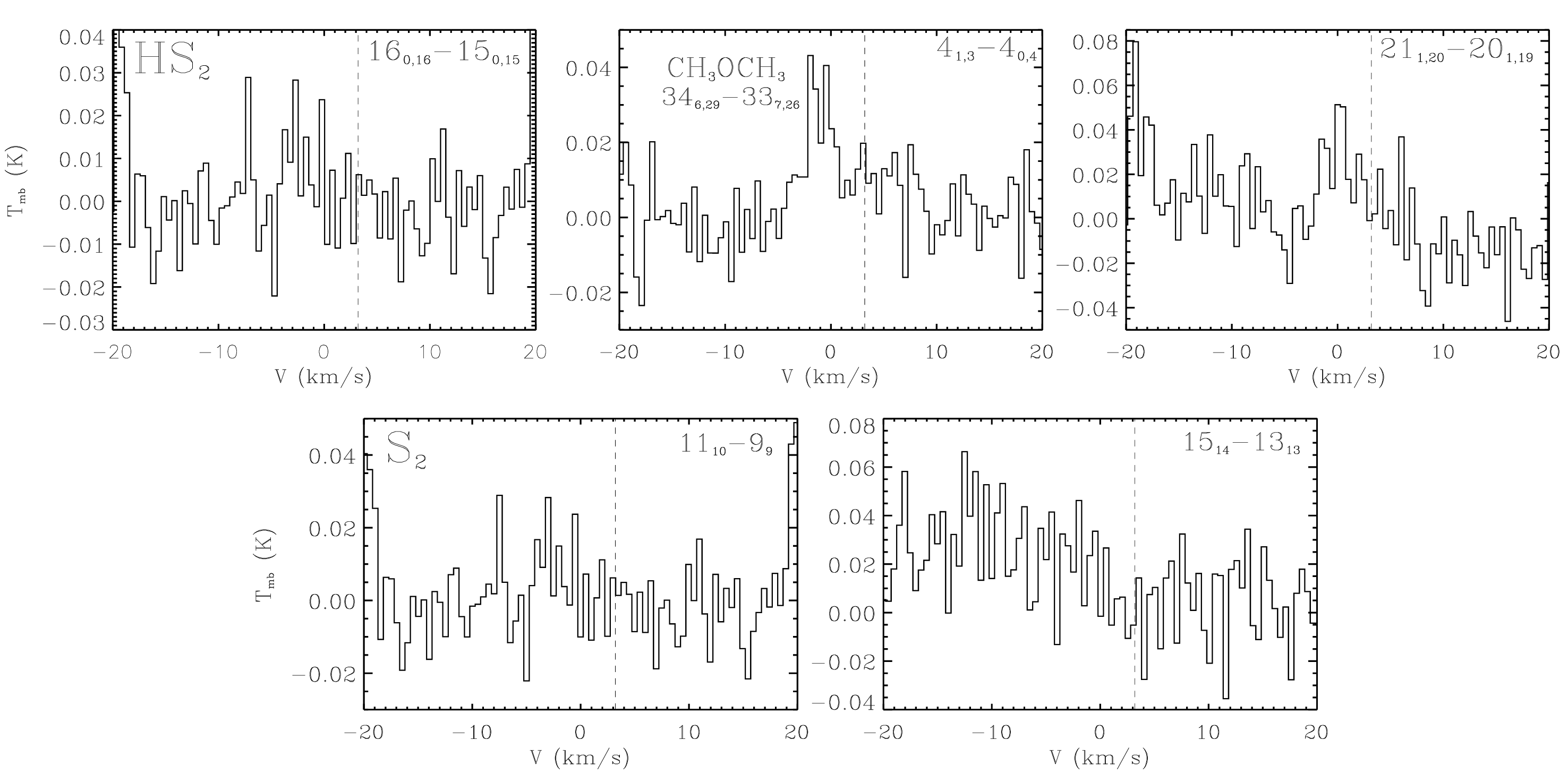}
\caption{Spectral windows where the HS$_{2}^{.}$ (top panels) and S$_{2}$ (bottom panels) transitions listed in Table \ref{lineas} should be 
detected. Vertical dashed lines mark a central radial velocity of v$_{\rm{lsr}}$ = 3.2 km s$^{-1}$, assumed for IRAS16293 A \citep{jorgensen11}.}
\label{hs2s2}
\end{figure*}

As mentioned above,  
we detected transitions from a collection of molecular species 
previously 
observed 
in other works 
such as C$^{17}$O, SO$_{2}$, HDCO, CH$_{3}$OH, CH$_{3}$CHO, CH$_{3}$OCHO, CH$_{3}$OCH$_{3}$, and others. 
Estimated column densities extracted from these emission lines are consistent with the column densities previously reported for these species 
\citep[see, e.g.,][]{cazaux03,caux11}.  
Some of these transitions are 
present 
in the spectral windows of Figures \ref{h2s2} and \ref{hs2s2}  
near the frequencies where the targeted transitions should be detected, and shortly described below. 

Fig. \ref{h2s2} shows one transition corresponding to CH$_{3}$CHO (13$_{2,12}$-14$_{1,14}$ panel) with E$_{\rm{up}}$ = 92.6 K, 
%
%
%
and one transition corresponding to HDCO (14$_{1,13}$-13$_{0,13}$ panel) with E$_{\rm{up}}$ = 56.3 K. 
The linewidth of the latter may indicate that HDCO is probing outflow material 
\citep[IRAS16293-2422 is embedded in the molecular cloud L1689N, which harbors multiple outflows;][]{wakelam04b}. 

In the case of CH$_{3}$OH, 
one transition with E$_{\rm{up}}$ = 203 K 
is detected in the 11$_{2,10}$-12$_{1,12}$ panel of Fig. \ref{h2s2}. 
The double-peaked structure of the line may indicate that the emission is also originated in outflowing gas. 
The double-peaked line detected in the 10$_{1,9}$-9$_{0,9}$ panel of Fig. \ref{h2s2} may correspond 
to the 27$_{3}$-27$_{2}$ CH$_{3}$OH transition. 
However, the energy of this transition may be too high, even for an outflow (E$_{\rm{up}}$ = 926.1 K). 

Four blended CH$_{3}$OCH$_{3}$ transitions with E$_{\rm{up}}$ = 38.2 K 
lead to three emission peaks 
in the 9$_{2,7}$-10$_{1,9}$ panel of Fig. \ref{h2s2}.  
They are possibly also blended with the 14$_{11,4}$-13$_{11,3}$ H$_{2}$CCO transition (E$_{\rm{up}}$ = 113.9 K) 
that should appear at a slightly higher frequency and therefore at more blue-shifted velocities. 
The weaker 30$_{4,26}$-30$_{3,27}$ CH$_{3}$OCH$_{3}$ transition 
is probably blended with the 23$_{13,10}$-22$_{13,9}$ CH$_{3}$OCHO transition 
in that spectral window (E$_{\rm{up}}$ = 433.7 K and 462.6 K, respectively). 
Other four blended transitions with high upper level energy (E$_{\rm{up}}$ = 593.9 K) 
corresponding to dimethyl ether 
are detected in the 4$_{1,3}$-4$_{0,4}$ panel of Fig. \ref{hs2s2}. 

Three more CH$_{3}$OCHO transitions are shown in Fig. \ref{h2s2}, 
one in the 9$_{2,7}$-10$_{1,9}$ panel (E$_{\rm{up}}$ = 173.2 K), 
and 
two blended 
transitions 
in the 5$_{2,4}$-6$_{1,6}$ panel (E$_{\rm{up}}$ = 277.8 K for both transitions). 

\subsection{Non-detection of H$_{2}$S$_{2}$, HS$_{2}^{.}$ or S$_{2}$}
\label{azufradas}

In Figures \ref{h2s2} and \ref{hs2s2}, we show the spectral windows where the transitions listed in Table \ref{lineas} should have been 
detected. The 6$_{1,5}$-5$_{0,5}$ panel of Fig. \ref{h2s2} presents some emission centered at v$_{\rm{lsr}}$ $\sim$ 3.2 km s$^{-1}$, which may 
correspond to the 6$_{1,5}$-5$_{0,5}$ transition of H$_{2}$S$_{2}$. However, the lack of detection of other transitions of this molecule 
prevents us from claiming this identification. In addition, the observed feature may present some contribution from the 18$_{5,14}$-17$_{5,13}$ 
CH$_{3}$OCHO transition (E$_{\rm{up}}$ = 305.1 K).  
Other possible lines of HCOCN and NCHCCO could be overlapping with this emission. However, their frequencies are typically shifted by $\sim$ 
0.5 MHz with respect to that of the H$_{2}$S$_{2}$ transition, which would imply an offset of $\sim$ 0.6 km s$^{-1}$ in the v$_{\rm{lsr}}$. 
The 14$_{11,4}$-14$_{10,5}$ transition of c-C$_{3}$D$_{2}$ (an isotopologue not reported in \citet{cazaux03} and \citet{caux11}) 
with E$_{\rm{up}}$ = 271.0 K 
could also contribute to the observed emission, since the uncertainty of the expected frequency is larger than 5 MHz. 

For HS$_2^{.}$ and S$_2$ (see Figure \ref{hs2s2}), there are no clear emission features arising at the v$_{\rm{lsr}}$ of the source, except for the 
4$_{1,3}$-4$_{0,4}$ panel. However, as mentioned in Section 4.1, the detected feature is likely associated with four blended transitions of 
dimethyl ether. 

From all this, we conclude that none of the transitions of H$_2$S$_2$, HS$_2^{.}$ or S$_2$ observed toward IRAS16293-2422 are detected. 
In Table \ref{lineas}, we provide the noise rms values over a linewidth of 5 km s$^{-1}$ 
in the spectral region around the frequency of each transition. The measured rms values range from $\sim$1.8 to 5.2$\,$mK, which provide 
stringent constraints on the upper limits to the molecular column densities and abundances of these species toward this low-mass warm core 
(see Section \ref{upper}).
  
%
%

\begin{table*}
\centering
\caption{Upper limits to the column densities and molecular abundances of H$_{2}$S$_{2}$, HS$_{2}^{.}$, and  S$_{2}$ estimated from 
the observations, and abundances calculated in laboratory experiments.}
\begin{tabular}{ccccc}
\hline
\hline
  & N$_{\rm{mol}}$ & N$_{\rm{mol}}$/N(H$_{2}$) & [N$_{\rm{mol}}$/N(H$_{2}$S)]$_{\rm{obs}}$ & [N$_{mol}$/N(H$_{2}$S)$_{i}$]$_{\rm{lab}}$ $^{(1)}$\\ 
Molecule & (cm$^{-2}$) & & (\%) & (\%)\\
\hline
H$_{2}$S$_{2}$ & $\leq$ 6.1 x 10$^{14}$ & $\leq$ 8.1 x 10$^{-9}$ & $\leq$ 1.5 & $\leq$ 25.0 $^{(2)}$\\ 
HS$_{2}^{.}$ & $\leq$ 8.0 x 10$^{14}$ & $\leq$ 1.1 x 10$^{-8}$ & $\leq$  2.0 & 15.0 \\
S$_{2}$ & $\leq$ 2.2 x 10$^{16}$ & $\leq$ 2.9 x 10$^{-7}$ & $\leq$   55.0 & ...$^{(3)}$ \\
\hline
\end{tabular}
\begin{list}{}
\item(1) After irradiation of pure H$_{2}$S ices with a total fluence of $\sim$ 1.5 x 10$^{18}$ photons cm$^{-2}$. 
Abundances are relative to the initial H$_{2}$S in the experiments. Derived from \citet{jimenez12}. 
\item(2) This upper limit is estimated considering that all the sulfur atoms not contained in H$_{2}$S or HS$_{2}^{.}$ molecules 
at the end of the experiments reported in \citet{jimenez12} are forming H$_{2}$S$_{2}$. 
\item(3) S$_{2}$ was not quantified since it was detected by mass spectrometry during warm-up of the UV-processed ice analogs in \citet{jimenez12}, 
but not by infrared spectroscopy. 
\end{list}
\label{densidades}
\end{table*}


\section{Upper limits to the column densities and abundances of H$_{2}$S$_{2}$, HS$_2^{.}$ and S$_2$}
\label{upper}

From the 1$\sigma$ rms values shown in Table \ref{lineas}, we 
have estimated the 1$\sigma$ upper limits to the column densities and abundances 
of H$_2$S$_2$, HS$_2^{.}$ and S$_2$ measured toward IRAS16293-2422. 
To this purpose, we have used the simple model mentioned in Section \ref{results}, assuming  
that the emission of these molecules is optically thin, that 
their line profiles have linewidths of 5 km s$^{-1}$, and that the excitation temperature of the gas is 100 K.  
The column densities of the three  
species 
were subsequently 
corrected from beam dilution using the beam filling factor \citep{maret11} 

\begin{equation}
\centering
\eta = \frac{\theta_{source}^{2}+\theta_{beam}^{2}}{\theta_{source}^{2}}
\end{equation}

\noindent and considering a source extension of 2$\arcsec$ in diameter \citep{wakelam04b}. 
In addition, the 1$\sigma$ upper limits for HS$_{2}^{.}$ are computed taken into account that the three pairs of lines cannot 
be resolved in our observations. 

The upper limits to the column densities of H$_2$S$_2$, HS$_{2}^{.}$ and S$_{2}$ are shown in the last column of Table \ref{lineas}. 
The most stringent values are $\leq$6.1$\times$10$^{14}$ $\,$cm$^{-2}$, $\leq$8.0$\times$10$^{14}$ $\,$cm$^{-2}$, and 
$\leq$2.2$\times$10$^{16}$ $\,$cm$^{-2}$ respectively (see also Table \ref{densidades}, column 2). 

The upper limits to the abundance of these molecules with respect to H$_2$ are derived by using an H$_{2}$ column density of N(H$_{2}$) = 7.5 x 
10$^{22}$ cm$^{-2}$ \citep{ceccarelli00} 
These upper limits are shown in column 3 of Table  \ref{densidades} and are $\leq$8.1$\times$10$^{-9}$ for H$_2$S$_2$, $\leq$1.1$\times$10$^{-8}$ 
for HS$_{2}^{.}$, and $\leq$2.9$\times$10$^{-7}$ for S$_{2}$. 

Since no molecule with a sulfur-sulfur bond has been detected so far in the interstellar medium, the estimated upper limits can be compared 
only to the abundances of species with just one sulfur atom.  
In this case, we have calculated the upper limits to the abundance of these species with respect to H$_2$S in the same source, 
which is thought to be the parent molecule of the targeted species.  
We have considered an H$_{2}$S column density of N(H$_{2}$S) = 4 x 10$^{16}$ cm$^{-2}$ as reported in \citet{wakelam04b}. 
The relative upper limits are shown in column 4 of Table \ref{densidades} and are $\leq$1.5\%, $\leq$2.0\% and $\leq$55.0\% 
for H$_2$S$_2$, HS$_{2}^{.}$ and S$_{2}$, respectively, 
which leads to a rough upper limit of about 1\% of S-S bonded species with respect to those with only one S atom. 

\section{Comparison with laboratory H$_{2}$S-bearing ice irradiation experiments}
\label{comparison}

The upper limits to the abundances relative to H$_{2}$S are compared to the values expected from the laboratory experiments peformed by 
\citet{jimenez11}, 
also shown in Table \ref{densidades} (column 5). The measured upper limits are factors of 7-17 lower than the abundances measured in the 
experiments. 

Differences between experimental and observed abundances could arise, on one hand, from the differences between the experimental simulations 
performed in the laboratory and the real processes taking place in the interstellar and circumstellar medium, 
leading to an overestimation of the expected abundances, as it is explained below. 
%
%
The laboratory simulations reported in \citet{jimenez11} used to derive the abundances shown in the fifth column of 
Table \ref{densidades} were carried out with pure H$_{2}$S ices. 
However, solid H$_{2}$S is expected to be found in a water-rich environment in the interstellar and circumstellar medium, 
along with other hydrogenated species \citep{allamandola99,jimenez11}.  
Since H$_{2}$S has not been firmly detected in interstellar or circumstellar ices, this scenario remains to be confirmed,  
but processing of solid H$_{2}$S could take place 
to a limited extent in multicomponent ices. 
In the dense interstellar medium, solid H$_{2}$S can be processed in dense cloud interiors by the secondary UV-field produced 
by excitation of H$_{2}$ molecules by cosmic rays, 
and/or by the radiation emitted by the central object in hot cores and hot corinos. 
During the experiments in \citet{jimenez12}, 
pure H$_{2}$S ice analogs experienced a total fluence of $\sim$ 1.5 x 10$^{18}$ UV-photons cm$^{-2}$,  
which is higher than the fluence of $\sim$ 3.2 x 10$^{17}$ photons cm$^{-2}$ that ice mantles are expected to experience 
in dense cloud interiors after 10$^{6}$ yr, assuming a flux value of 10$^{4}$ photons cm$^{-2}$ s$^{-1}$ for the secondary UV-field 
\citep[and ref. therein]{shen04}. 

On the other hand,  
the region affected by the X-ray radiation emitted by a protostar like IRAS16293 A, with a typical extension of $\sim$ 10 AU 
\citep{ciaravella11}, 
is smaller than the region where the sulfur-containing species are expected to be detected in these observations 
\citep[with an extension of $\sim$ 230 AU for an assumed source size of 2$\arcsec$ in diameter;][]{wakelam04b}. 
Therefore, the upper limits shown in the third and fourth columns of Table \ref{densidades} could be affected by beam dilution, 
(and, therefore, be underestimated)  
if the sulfur-bearing molecules were confined to the region affected by the X-ray emitted by the protostars in IRAS16293-2422. 

Finally, H$_{2}$S$_{2}$ and HS$_{2}^{.}$, although efficiently produced in ices, may undergo an active gas-phase chemistry once desorbed 
from dust grains. This could lower the gas-phase abundances of these species to undetectable levels. 
This possibility is further explored in Section \ref{model}. 

\begin{figure*}
\centering
\includegraphics[width=18cm]{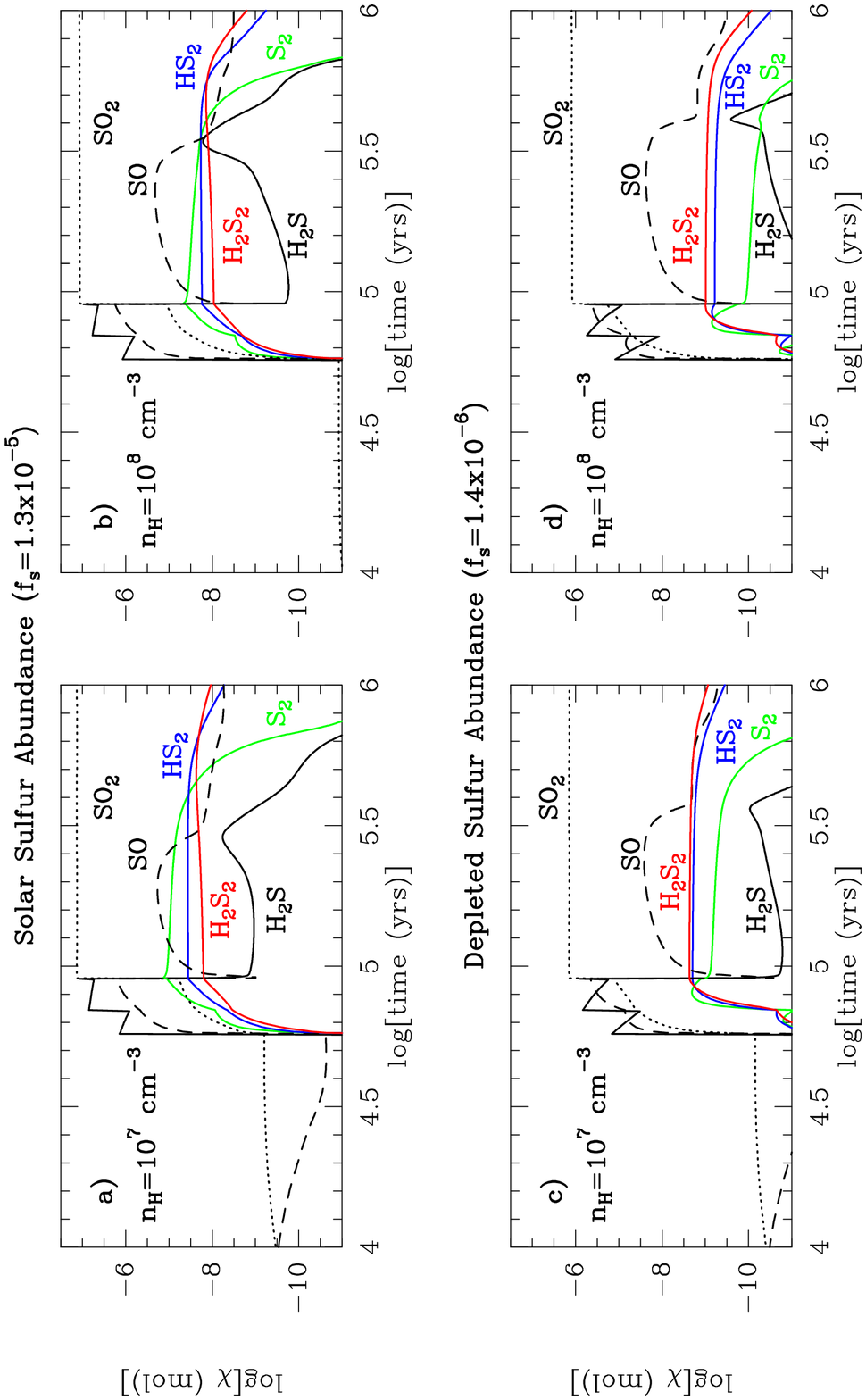}
\caption{Time-dependent evolution of abundances of selected sulfur-bearing species in IRAS16293 according to the UCL\_CHEM code for models 
(a), (b), (c), and (d), whose most important parameters are shown in Table \ref{models}. 
Colored lines correspond to the targeted species. Black solid line is for H$_{2}$S. Black dashed and dotted 
lines correspond to SO and SO$_{2}$, respectively.}
\label{modelsfig}
\end{figure*}

\section{Chemical modeling of H$_{2}$S$_{2}$, HS$_{2}^{.}$ and S$_{2}$ in IRAS16293-2422}
\label{model}
In a recent paper, \citet{woods15} have incorporated the experimental data of \citet{garozzo10} on the production of the S-bearing species 
H$_{2}$S$_{2}$, HS$_{2}^{.}$ and S$_{2}$ in ices, into the UCL\_CHEM chemical code \citep{viti99,viti04}. 
In these experiments, \citet{garozzo10} bombarded a CO:H$_{2}$S = 10:1 mixture sample with 200 keV protons simulating the impact of cosmic rays 
on interstellar ice analogs. Although 
the irradiation source is different 
from 
the ones used in 
\citet{jimenez11} and \citet{jimenez14}, 
we note that the measured 
production of H$_{2}$S$_{2}$ in the ices at low fluences is about 30\%, i.e., similar within 5\% to the one obtained in the experiments 
with UV-photon and X-ray irradiation (see Table \ref{densidades}). 
Since we are interested in the chemistry of these sulfur-bearing species in the gas phase 
after thermal desorption, a small difference in the H$_{2}$S$_{2}$ production 
by 5\% 
is not expected to have a noticeable effect in the final results. 
We therefore made use of this new chemical network to test whether gas-phase chemistry is responsible for the lack of detection of 
H$_{2}$S$_{2}$, HS$_{2}^{.}$ and S$_{2}$ toward IRAS16293. 

The UCL\_CHEM gas-grain chemical code is a two-phase model. 
In Phase I, this  code  simulates  the  free-fall  gravitational  collapse  of a cloud where the mantles of dust grains form self-consistently 
via freeze-out reactions. 
In our modeling, we have considered two groups of freeze-out reactions for the sulfur species: one group 
where molecules can either freeze onto grains as themselves, or undergo hydrogenation/dehydrogenation 
reactions in the ices; 
and a second group  
where species freeze onto grains only as themselves
(see Table \ref{models} for the freeze-out reactions considered in the models). 
The formation of the  initial core  occurs from a diffuse medium with H$_{2}$ densities of $\sim$100 cm$^{-3}$ and the collapse stops when 
the final density is reached in the model. 
The UCL\_CHEM is run for two final densities of 10$^{7}$ cm$^{-3}$ and 10$^{8}$ cm$^{-3}$, covering the range of H$_{2}$ gas densities typically 
assumed for low-mass warm cores \citep[see, e.g.,][]{awad10}. 
Finally, in order to test the effects of the presence of a sulfur residue in dust grains in the form of S$_{8}$ or of any other type of 
aggregate \citep[see][]{jimenez11}, we also considered two models with different initial sulfur abundances (see Table \ref{models}): 
one assuming the solar sulfur abundance \citep[f$_{s}$=1.318$\times$10$^{-5}$;][]{asplund09} and a second model considering a depleted 
abundance of sulfur \citep[f$_{s}$=1.4$\times$10$^{-6}$;][]{sofia94}. 
%
In Phase II, UCL\_CHEM calculates the time-dependent evolution of the chemistry of gas and dust once stellar activity is present. 
As explained in \citet{viti04}, molecular species are thermally evaporated after the turning on of the protostar in different temperature bands. 
Here, we follow the classification proposed by \citet{woods15} where H$_2$S, HS and H$_2$S$_2$ are considered as \textit{intermediate} species 
evaporating at temperatures between 80 K - 90 K, and where HS$_{2}^{.}$ is a \textit{reactive} molecule that co-desorbs with water at around 100 K. 
The maximum temperature assumed for IRAS16293 in Phase II is 100 K \citep{ceccarelli00}. 
In our models the temperature increases monotonically following the prescription of \citet{viti04}. 
We also assume a standard cosmic ray ionisation rate ($\zeta$=1.3$\times$10$^{-17}$ s$^{-1}$) for all models. 

\begin{table*}
\centering
\caption{Models run for the chemistry of H$_{2}$S$_{2}$, HS$_{2}^{.}$ and S$_{2}$ toward IRAS16293}
\begin{tabular}{ccccc}
\hline
\hline
&&&Freeze-out&\\
Model&n$_{H_{2}}$ (cm$^{-3}$)&f$_{s}$&reactions group&Reactions$^{(1)}$\\
\hline
&&&&50\% HS$_{2}$ $\rightarrow$ mHS$_{2}$\\
&&&&50\% HS$_{2}$ $\rightarrow$ mH$_{2}$S$_{2}$\\
a)&1$\times$10$^{7}$&1.3$\times$10$^{-5}$&1&50\% H$_{2}$S$_{2}$ $\rightarrow$ mH$_{2}$S$_{2}$\\
&&&&50\% H$_{2}$S$_{2}$ $\rightarrow$ mHS$_{2}$\\
b)&1$\times$10$^{8}$&1.3$\times$10$^{-5}$&1&50\% S $\rightarrow$ mHS\\
&&&&50\% S $\rightarrow$ mH$_{2}$S\\
&&&&0\% S $\rightarrow$ mS\\
\hline
&&&&50\% HS$_{2}$ $\rightarrow$ mHS$_{2}$\\
&&&&50\% HS$_{2}$ $\rightarrow$ mH$_{2}$S$_{2}$\\
c)&1$\times$10$^{7}$&1.4$\times$10$^{-6}$&1&50\% H$_{2}$S$_{2}$ $\rightarrow$ mH$_{2}$S$_{2}$\\
&&&&50\% H$_{2}$S$_{2}$ $\rightarrow$ mHS$_{2}$\\
d)&1$\times$10$^{8}$&1.4$\times$10$^{-6}$&1&50\% S $\rightarrow$ mHS\\
&&&&50\% S $\rightarrow$ mH$_{2}$S\\
&&&&0\% S $\rightarrow$ mS\\
\hline
&&&&100\% HS$_{2}$ $\rightarrow$ mHS$_{2}$\\
&&&&0\% HS$_{2}$ $\rightarrow$ mH$_{2}$S$_{2}$\\
e)&1$\times$10$^{7}$&1.3$\times$10$^{-5}$&2&100\% H$_{2}$S$_{2}$ $\rightarrow$ mH$_{2}$S$_{2}$\\
&&&&0\% H$_{2}$S$_{2}$ $\rightarrow$ mHS$_{2}$\\
f)&1$\times$10$^{8}$&1.3$\times$10$^{-5}$&2&0\% S $\rightarrow$ mHS\\
&&&&0\% S $\rightarrow$ mH$_{2}$S\\
&&&&100\% S $\rightarrow$ mS\\
\hline
\end{tabular}
\begin{list}{}
\item$^{(1)}$mH$_{2}$S$_{2}$, mHS$_{2}$, mH$_{2}$S, and mS refers to species frozen onto the mantles of dust grains.
\end{list}
\label{models}
\end{table*}

In Figure \ref{modelsfig}, we report the abundances of all major species involved in the sulfur chemistry toward IRAS16293. 
Consistently with the results of \citet{woods15}, 
the level of hydrogenation of the ice species in Phase I does not significantly affect 
the abundances of these species in Phase II. Gas phase reactions after ice desorption clearly dominate the chemistry of S-bearing species. 
The abundances of H$_{2}$S$_{2}$, HS$_{2}^{.}$ and S$_{2}$ in the mantles at the end of Phase I 
indeed differ by only 
factors of a few between the two freeze-out cases considered in our models. 
Therefore, in Figure \ref{modelsfig} we only show the modeling results for the cases where molecules freeze out both as themselves or undergoing 
hydrogenation/dehydrogenation in the ices (see Group 1 reactions in Table \ref{models}).  
%
Differences are found between models with different H$_{2}$ gas densities (see, e.g., models (a) and (b)).
In particular, the abundances of H$_{2}$S$_{2}$, HS$_{2}^{.}$ and S$_{2}$ are factors 1.7 - 6.7 higher in the models with n$_{H_{2}}$=10$^{7}$ cm$^{-3}$ 
than for H$_{2}$ gas densities of 10$^{8}$ cm$^{-3}$. One explanation is that more H$_{2}$S is formed on grain surfaces at higher densities, 
which in turn is converted into SO$_{2}$ via the surface reaction H$_{2}$S + 2O $\rightarrow$ SO$_{2}$ + H$_{2}$ \citep[see Table 3 in][]{woods15}. 
Since H$_{2}$S preferentially goes into SO$_{2}$ at high densities, there is a smaller amount of this molecule available in the ices for the 
production of H$_{2}$S$_{2}$, HS$_{2}^{.}$ and S$_{2}$, leading to lower abundances of the latter species. 
As expected, Figure \ref{modelsfig} also shows that a depleted initial abundance of atomic sulfur yields lower abundances in Phase II 
for all S-bearing species considered in our modeling.  

In order to compare the model results with our upper limits of Section \ref{upper}, we need to inspect the abundances of 
H$_{2}$S$_{2}$, HS$_{2}^{.}$ and S$_{2}$ at time-scales $\sim$10$^{5}$ yrs inferred for the IRAS16293 low-mass warm core 
\citep[see][]{andre93}. 
The predicted abundances of these species for models (a), (b), (c), and (d) are shown in Table \ref{abundancias} along with the upper 
limits derived from our observations. 
For the models with a sulfur solar abundance (models (a) and (b)), higher H$_{2}$ densities of 10$^{8}$ cm$^{-3}$ fit better the 
observations (see Table \ref{abundancias}). 
However, for models (c) and (d) where the initial sulfur abundance is depleted, we find that the predicted abundances of  
H$_{2}$S$_{2}$, HS$_{2}^{.}$ and S$_{2}$ lie within the upper limits measured toward IRAS16293. 
Therefore, even in the absence of a sulfur residue, our modeling shows that gas-phase chemistry can explain the lack of detection of 
H$_{2}$S$_{2}$, HS$_{2}^{.}$ and S$_{2}$ toward this source. 

\begin{table*}
\centering
\caption{Comparison of the modelled H$_{2}$S$_{2}$, HS$_{2}^{.}$ and S$_{2}$ abundances with the upper limits derived from the observations.}
\begin{tabular}{cccccc}
\hline
\hline
&N$_{\rm{mol}}$ (cm$^{-2}$)&N$_{\rm{mol}}$ (cm$^{-2}$)&N$_{\rm{mol}}$ (cm$^{-2}$)&N$_{\rm{mol}}$ (cm$^{-2}$)\\
Molecule&Model (a)&Model (b)&Model (c)&Model (d)&Observations\\
\hline
H$_{2}$S$_{2}$&1.6$\times$10$^{-8}$&9.3$\times$10$^{-9}$&2.3$\times$10$^{-9}$&1.0$\times$10$^{-9}$&$\leq$8.1$\times$10$^{-9}$\\
HS$_{2}^{.}$&3.6$\times$10$^{-8}$&1.7$\times$10$^{-8}$&2.0$\times$10$^{-9}$&6.2$\times$10$^{-10}$&$\leq$1.1$\times$10$^{-8}$\\
S$_{2}$&1.1$\times$10$^{-7}$&3.7$\times$10$^{-8}$&7.4$\times$10$^{-10}$&1.1$\times$10$^{-10}$&$\leq$2.9$\times$10$^{-7}$\\
\hline
\end{tabular}
\label{abundancias}
\end{table*}

\section{Conclusions}
\label{conclusiones}
A fraction of the missing sulfur in the interstellar medium is thought to be locked in the ice mantles on dust grains. 
Interstellar ice mantles are energetically processed, leading to chemical and structural changes. 
Previous laboratory experiments simulating the irradiation of H$_{2}$S-containing ice analogs 
under astrophysically relevant conditions lead to the production 
of H$_{2}$S$_{2}$, HS$_{2}^{.}$, and  S$_{2}$ among other species. 
These molecules would be subsequently released to the gas phase by non-thermal desorption in dense clouds, 
and also by thermal desorption in regions around YSOs. 
We have presented the first single-dish observations targeting 
a wide selection of rotational transitions predicted for the S-S bonded molecules  
H$_{2}$S$_{2}$, HS$_{2}^{.}$, and  S$_{2}$ 
toward the low-mass warm core IRAS16293-2422, 
a region with an active chemistry and a high H$_{2}$S gas-phase abundance. 
As a result of this work, the predicted rotational spectrum of H$_{2}$S$_{2}$ was made available as a new entry in the CDMS catalog.  
Although none of the species were firmly detected, we have estimated an upper limit to their molecular abundances of 
$\leq$8.1 x 10$^{-9}$, $\leq$1.1 x 10$^{-8}$, and $\leq$2.9 x 10$^{-7}$ relative to H$_{2}$ 
or $\leq$1.5 x 10$^{-2}$, $\leq$2.0 x 10$^{-2}$, and $\leq$5.5 x 10$^{-1}$ relative to H$_{2}$S, respectively.   
These abundances are 
therefore up to two orders of magnitude lower than the observed abundance of H$_{2}$S in the same source, 
which is thought to be the parent molecule of species with a disulfide (S-S) bond.  
The estimated upper limits are one order of magnitude lower than the abundances found in the experimental simulations.  
This could be the result of an 
underestimation of the measured upper limits caused by beam dilution of their emitting regions,   
or an overestimation of the expected abundances from the laboratory experiments.  
Although experimental simulations try to mimick interstellar conditions, differences between the ice analogs and the interstellar ice mantles, 
the densities in the UHV chamber and the interstellar and circumstellar regions, and the processing time of the samples are inevitable. 
On the other hand, subsequent gas-phase chemistry after desorption could have reduced the gas-phase abundances of these species 
to undetectable levels in our observations. 
Future high-angular resolution observations will establish whether these species are 
truly depleted in the gas phase or whether their emitting regions are largely diluted in our single-dish observations. 

\begin{acknowledgements}
This research was financed by the Spanish MINECO under project AYA2011-29375. 
R.M.-D. benefited from a FPI grant from Spanish MINECO. 
I.J.-S. acknowledges the funding received from the People Programme (Marie Curie Actions) of the European Union’s Seventh Framework Programme 
(FP7/2007-2013) under REA grant agreement PIIF-GA-2011-301538, and from the STFC through an Ernest Rutherford Fellowship (proposal number 
ST/L004801/1).
\end{acknowledgements}


\end{document}